# Describing and Modeling of Video-on-Demand Service with the Usage of Multi-Layer Graph

Dmitry Ageyev, Artem Ignatenko

*Abstract* – **Considered in this paper is the method of describing of telecommunications systems providing VoD service using multi-layer graph. The paper describes the relations between the structural elements at each hierarchical level of the multi-layer graph. Transfer of video is a resource consuming task, and it requires an optimal configuration of the studied system. The usage of the multi-layer graph makes it possible to consider the telecommunication system as a whole and avoid falling in the local optimums when solving optimization problems.**

*Keywords* – **Video-on-Demand, Multi-Layer graph, Network design, Topology synthesis.**

## I. INTRODUCTION

The Video-on-demand (VoD) service is an integral part of the modern multi-service network and can be considered as a subset of IPTV service. System for delivering VoD service can be described as a set of servers that host video content. These servers are interconnected in a certain way (either directly or through intermediate nodes) and they also must have connections with the subscribers through subscriber access nodes.

Telecommunication systems are constructed hierarchically, on the basis of overlay networks, i.e. each lower level of the network provides transparent transfer of the network flow over the upper level.

Thus, it is expedient to represent such networks as an ordered set of graphs, each of which describes the topology of a network at a particular hierarchical level. So, because of the complexity of such systems, it has been proposed to use multi-layer graphs for their description, as shown at [1, 2].

## II. MULTI-LAYER GRAPH DESCRIPTION

The system for delivering VoD can be described as a set V of servers that host video. Thus, we define all stakeholders providing VoD services as:

$A = \{a_i\}$ - set of subscribers of VoD service;
$VS = \{vs_i\}$ - set of video-servers;
$X = \{x_i\}$ - set of intermediate nodes.
$NA = \{na_i\}$ - set of subscriber access nodes;

Dmitry Ageyev - Kharkiv National University of Radioelectroniks, Lenina Av., 14, Kharkov, 61166, UKRAINE, E-mail:dm@ageyev.in.ua

Artem Ignatenko – Kharkov National University of Radioelectronics, Lenina str., 14, Kharkiv, 61166, UKRAINE, E-mail:sanitarium@ukr.net

Desription of a telecommunication system as a multi-layer graph is done in accordance with the following method [3]:
1. Distingush a set of levels in the telecommunication system being modelled.
2. Describe the topology of each level with the help of a classic graph
3. Distinguish logic, functional and physical connections between the objects of different levels and describe them using the graphs.
4. Assign a set of characteristics to the corresponding parameters of the objects and inter-object relations interesting for the modelling, to the edges and vertices of the graph

The application of this method makes it possible to receive a multi-layer graph that has the structure as shown on Fig. 1.
V is a set of vertices of the multi-layer graph G (see Fig. 1).

$$G = \{\Gamma^1, \Gamma^2, \Gamma^3\},$$
$$V = A \cup VS \cup NA \cup X,$$
$$A \cap VS = \varnothing, \quad A \cap NA = \varnothing,$$
$$A \cap X = \varnothing, \quad VS \cap NA = \varnothing.$$

Now we must describe each level of G:
− cooperation between the service and subscribers ($\Gamma^3$);
− cooperation between the subscribers and the servers, and the servers with each other ($\Gamma^2$);
− entire physical topology of the network ($\Gamma^1$).

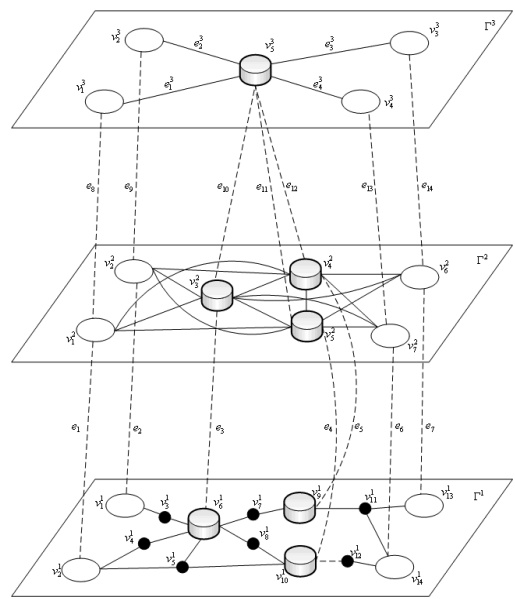

Fig.1 Description of video-on-demand service with a multi-layer graph

On the level of interaction of the subscribers with the service, the graph will have the "star" topology since all the subscribers interact with the service.

On the level of interaction of the subscribers with the servers and the servers between themselves, the topology will be almost mesh topology, because it is envisaged that each VoD server can interact with the other servers, and each subscriber can receive content from each server. This being said, the subscribers have no connections between themselves, so there will be no edges between the graph vertices that correspond to the subscribers. Finally, the bottom level describes the physical interaction between the VoD servers, subscribers and access nodes between themselves.

The concept of the multi-layer graph implies the modeling of overlay networks. Consequently, each functional unit on each level has its corresponding counterpart on the other adjacent level $E=\{e_1,e_2,\ldots,e_{14}\}$. Such correspondence on the multi-layer graph is expressed by the edge joining the functional units (vertices) corresponding to each other on the adjacent levels of the multi-layer graph.

Now it is possible to describe the flow model on the multi-layer graph [3]. The rule of flow conservation must be respected.

Generally, the flow conservation law can be formulated in three statements.

− Amount of flow transferred between the pair of the interacting source-destination nodes along the chosen route is always unchanged.

− Sum of flows transferred by different routes between the pair of the interacting source-destination nodes equals to the amount of requirements emerging in the source node, and equals to the amount of requirements processed in the destination node.

− Sum of flows incoming to the node equals the sum of flows outgoing from the node if the node functions only as a transit node; or it differs by the amount equaling to the difference between the amounts of flows for which it is a source or a destination.

The next step is to define the constraints for each level of the multi-layer graph. General constraints for the network graph should be complimented by the constraints characteristic for the VoD systems. In this way, generally, the subscribers cannot function as transit nodes, i.e. they cannot generate traffic in the framework of the system being considered and transfer it to the other nodes.

The lowest level subgraph $\Gamma^1$ is a redundant graph including all intermediate nodes and access points. Its redundancy enables us to solve the problem of network topology synthesis on the redundant graph. Linear and linear-integer programming methods may be used here. This approach is known as node-link formulation of the topology synthesis problem and is described in [4].

Occupied bandwidth minimum can be used as the optimality criterion for the links used for the traffic transmission.

Conclusion

VoD service sets high requirements towards the parameters of the network providing the service. In order to meet the QoS requirements for VoD service, minimize the cost of equipment or maximize the income of the telecommunication services provider, it is necessary to solve the optimization problem.

Consequent solving of the design problem for each of the network levels separately, when the results of the design on one level become the source data for the remaining network levels, does not consider the system as a whole, but only provides an optimal result for each sublevel, which can lead to local optimums but not optimally solve the problem as a whole. This drawback can be eliminated by way of applying the mathematical model of multilayer graph during the planning of the VoD services systems.

References

[1] D.V. Ageyev, "Modelirovanie sovremennykh telekommunikatsionnykh sistem mnogosloinymi grafami" [Simulation of modern telecommunication systems with multi-layer graphs usage], *Problemi telekomunìkacìj, no1(1), pp.* 23 – 34, 2010, http://pt.journal.kh.ua/ 2010/1/1/ 101_ageyev_simulation.htm

[2] D.V. Ageyev "Metodika opisaniya struktury sovremennykh telekommunikatsionnykh sistem s ispol zovaniem mnogosloinykh grafov" [The method of describing the structure of modern telecommunication systems using multi-layer graph], *Eastern European Journal of Enterprise Technolopgies*, No 6/4 (48), pp. 56-59, 2010.

[3] D.V. Ageyev,, "Metod proektirovaniya telekommunikatsionnykh sistem s ispol zovaniem potokovoi modeli dlya mnogosloinogo grafa" [Method of designing telecommunication systems using flow model for multi-layer graph], *Problemi telekomunìkacìj,* No2(2), pp. 7 – 22, 2010, http://pt.journal.kh.ua/2010/2/2/ 102_ageyev_layer.htm.

[4] Micha Pióro, "Deepankar Medhi Routing, Flow, and Capacity Design in Communication and Computer Networks". Elsevier, 2004, 765 p.